\documentclass[11pt,pdftex]{article}

\usepackage{graphicx}
\usepackage{amsfonts, amsmath, amssymb, mathrsfs}
\usepackage{setspace}
\usepackage[utf8]{inputenc}
\usepackage{fullpage}

\newcommand{\be}{\begin{equation}} 
\newcommand{\ee}{\end{equation}}
\newcommand{\quadsp}{\;\;\;\;}

\newcommand{\half}{{\textstyle \frac{1}{2}}}
\newcommand{\tinyhalf}{{\scriptstyle \frac{1}{2}}}

\newcommand{\goesto}{\rightarrow}
\newcommand{\sech}{\,{\rm sech}\,}
\newcommand{\sltwor}{{\ensuremath {\rm SL(2,\mathbb{R})}} }
\newcommand{\sltwoc}{{\ensuremath {\rm SL(2,\mathbb{C})}} }
\newcommand{\sutwo}{{\ensuremath {\rm SU(2)} }}
\renewcommand{\Re}{\,{\rm Re}\,}

\usepackage[colorlinks=true,citecolor=blue]{hyperref}
\hypersetup{
pdfauthor=Bruno Carneiro da Cunha,
pdftitle=Tachyon Moduli and Closed Strings,
plainpages=false}

\begin{document}
\begin{titlepage}
\renewcommand{\thepage}{title}
\today

\vspace{3cm}
\centerline{ \Large \bf A Note on Tachyon Moduli and Closed Strings}

\vspace{2cm}

\centerline{ \normalsize Bruno Carneiro da
Cunha\footnote{bcunha@df.ufpe.br}}
\vspace{.5cm}
\centerline{\sl Helsinki Institute of Physics}
\centerline{\sl P.O. Box 64, FIN-00014 University of Helsinki, Finland}
\vspace{.3cm}
\centerline{and}
\vspace{.3cm}
\centerline{\sl Departamento de Física, Universidade Federal de
  Pernambuco\footnote{Permanent address}}
\centerline{\sl CEP 53901-970, Recife, Pernambuco, Brazil}
\vspace{1cm}

\begin{abstract}
The collective behavior of the $\sltwor$ covariant brane states of
non-critical $c=1$ string theory, found in a previous work, is studied
in the Fermi liquid approximation. It is found that such states
mimick the coset WZW model, whereas only by further restrictions one
recovers the double-scaling limit which was purported to be equivalent to
closed string models. Another limit is proposed, inspired by the
tachyon condensation ideas, where the spectrum is the same of
two-dimensional string theory. We close by noting some strange
connections between vacuum states of the theory in their different
interpretations.

\end{abstract}

\end{titlepage}

\newpage

\setcounter{footnote}{0}

\section{Some Remarks on Tachyon Condensation and 2-d 
String Theory}

Two dimensional models of strings have always been helpful to describe
various new phenomena in the theory. Discrete states, non-perturbative
models and the open-closed string duality being only a few. There is
however an inherent danger in taking symplified models given that the
algebraic structures tend to reduce to the same form, even though they
may arise from physically distinct properties.

The case at hand are the algebras related to $\sltwor$. The group is
so ubiquitous in physics that sometimes an interpretation of its
appearance becomes a daunting task. For instance, the relation of the
loop algebra of $\sltwor$ and $W_\infty$ has for long been known in the
context of two-dimensional gravity. The latter algebra has a multitude
of realizations, like for instance as the moments of fixed spin of
${\rm U(1)}$ currents in the two-dimensional free boson
field \cite{Ginsparg:1993is}. Applications of the algebra structure to
the double scaling limit of matrix models only furthers the problem
given its robustness with respect to the choice of potentials. There
is then the real danger that the lessons one takes from the symplified
models are properties not based on principles that can be applied in more
realistic theories.

In this letter we investigate the relationship between the  collective
behavior of unstable D0 branes and closed string backgrounds in two
dimensional string theory. We will start by arguing that the
collective behaviour is nothing more than the field theory of on-shell
boundary states of a single boson, which with few extra assumptions
can be shown to be equivalent to a gauged $\sltwor$ WZW model. The
latter will be interpreted as Dijkgraaf {\it et al.}
\cite{Dijkgraaf:1991ba}, as an exact background for closed
strings. The exercise shows a clear way of relating open string
collective states, in a suitable limit, to closed string
backgrounds. This is carried out in the spirit of
\cite{Karczmarek:2003pv}, among others.  

The construction is in way very similar to the one discussed in
\cite{Alekseev:1989npb}. This fact permits us to give a ``moduli
space'' interpretation of the appearance of the $\sltwor$ symmetry
discussed in \cite{daCunha:2004dh}, and also of its gauging. The
semiclassical limit is given by a dense Fermi fluid of D0 branes, and
we view it in light of the symmetry. We show how the model recovers
the double scaling limit, by taking excitations in a suitable limit of
a sector with fixed value of the quadratic Casimir of $\sltwor$. We
also show that the low-lying spectrum of the same sector does have an
equivalence to the spectrum of closed strings
\cite{Klebanov:1991qa}. We conclude by remarking a strange similarity
between the vacuum state of the coset model, with its
``gravitational'' or ``closed string'' characteristic and the vacuum
state of the bosonized collective field of branes, which can be
thought of as a scalar field in de Sitter spacetime.

\section{Single Particle Configuration Space}

It has long been known  that the boundary states
of a free boson at the self-dual radius have a $\sutwo$ group
structure (see, for instance, \cite{Callan:1994ub}). They can be
reinterpreted as the conformally invariant states of a brane on 
which a string with coordinate given by the single boson ends. The
crucial ingredient to the group structure are the generators:
\be
J^\pm = \oint \frac{dz}{2\pi i}e^{\pm i\sqrt{2}X(z)},\quadsp
J^3=\oint \frac{dz}{2\pi i}\frac{1}{\sqrt{2}}i\partial X(z)
\label{currents} 
\ee
which can be shown to have a well-defined action even at infinite
radius. The algebra gives a correspondence between 
on-shell boundary states for the Euclidean theory and $\sutwo$ group
elements, at least at the self-dual radius. To wit, they can all be
seen as an $\sutwo$ rotation of the Dirichlet state: 
\be
|B\rangle = e^{i\theta_a J^a}|D\rangle \label{bdrystate1},
\ee
with for instance the Neumann state being given by $\theta_a
J^a=\theta_1J^1+\theta_2J^2+\theta_3J^3=-\pi J^1$ in 
the notation of \cite{Callan:1994ub}. Because of the state-operator
correspondence, the lesson is that one also has a group
parametrization of the conformally invariant boundary interactions in this
model. This in turn gives a mapping between the zero mode of the
boson $x=X(0)$ and the value of the tachyon (boundary) interaction
$T(x)$ and ${\rm SU(2)}$:
\be
T(x)=\theta_+ e^{i (x-\theta_3)/\sqrt{2}}+ \theta_-
e^{-i(x-\theta_3)/\sqrt{2}},\label{tachyonmap}
\ee
which associates a line of $\sutwo$ elements to a given value of $x$ and
$T$, since one has the symmetry 
\be
\theta_\pm\goesto \theta_\pm
e^{\pm\alpha/\sqrt{2}}, \quadsp\theta_3 \goesto \theta_3 +
\alpha.\label{symmetry1} 
\ee 
This symmetry will be crucial in the main course of this paper. Its
space-time interpretation the two-dimensional gauge field,
whose effect for a single brane is just to trivialize the action of
the axial subgroup
\be
g\goesto hgh,\quadsp h=e^{i\alpha J_3}.\label{symmetry2}
\ee
In the following we will perform another abuse of the notation and call
$J_3$ as the operator that translates $x$. The natural parametrization
of $\sutwo$ is then
\be
g=e^{i(\phi+\alpha)J_3}e^{i\theta J_1}e^{-i(\phi-\alpha)J_3}.\label{euler}
\ee

The point of view taken in this paper is that the local algebra ${\rm
  su(2)}$ does provide a powerful tool for analysis of the
condensation process. The basis of the argument will be outlined in
the following paragraphs and sections. For now it is worth stressing
the proposal is {\it not} the notorious exact correspondence between the
free boson and the ${\rm SU(2)}$ WZW model at $k=1$. The construction
will be inspired by the McGreevy-Verlinde's \cite{McGreevy:2003kb}
``fluid of branes'' and as such we will be only interested in the zero
modes of the embedding coordinates $\{\phi,\theta\}$, instead of the
whole spetrum of the boson. Also, it is known
that, while the $\sutwo$ current (local) structure is pervasive, the
global structure of the moduli space can be quite different
\cite{Recknagel:1998ih}, with global identifications,
decompactifications and especial discrete states showing up depending
on which orbifold of the uncompactified boson one takes. These effects
can be accounted for by a suitable restriction on the allowed
representations of the algebra. All of this considered, we will have
the local structure of the boson at self-dual radius as paradigm: in
this case the configuration space is $\sutwo\simeq S^3$, which for the
single particle is reduced to $S^2$ as the effect of fixing the
symmetry (\ref{symmetry2}). 

As Sen argued in a series of papers,
(\cite{Sen:2002in,Garousi:2000tr}, see \cite{Sen:2004nf} for a 
review), the naive Wick 
rotation of the boson does provide a realistic picture of the tachyon
condensation process in two dimensions, where oscillator modes are
supressed by the BRST constraint. As far as the configuration space is
concerned, the Wick rotation brings the group structure to
$\sltwor$. This fact was anticipated by Gaberdiel {\it et al.},
\cite{Gaberdiel:2001zq}, where it is argued that if one does not
bother about unitarity, the moduli space of the boundary states at the
self-dual radius is just the complexified algebra generated by the
currents (\ref{currents}), $\sltwoc$. Some of the states would have a
strange interpretation in string theory: branes at imaginary
positions, for instance. With the finding that these do actually
correspond to closed string backgrounds \cite{Gaiotto:2003rm}, one is
tempted to take the assumption that the Wick rotated boson's conformal
configuration space is in fact just a truncation of the same
$\sltwoc$, with a reality condition suitable to study time-like
bosons. For the Euclidianized theory, {\it i.e.}, a space-like boson,
the truncation would yield the $\sutwo$ current structure reviewed
above. For the Lorentzian theory, {\it i. e.}, a time-like boson, the
truncation yields $\sltwor$ currents. In the latter case, the work of
\cite{Gaberdiel:2001zq} raised suspicions about whether the generic
states constructed in the Lorentzian case allow for a sensible Hilbert
space structure. We will have more to say about this below.

In the Euclidian theory, the global structure in the self-dual point
can be determined from the local algebra and properties of Ishibashii
states. When one studies the behavior of the states under generic
$\sutwo$ transformations, one begins with the Neumann state
$|N\rangle$ and applies to it finite $\sutwo$ transformations. By
considering the actual, effective, value of the tachyon that relates
the Dirichlet to the Neumann state in (\ref{bdrystate1}), one sees
that it is actually renormalized to a finite value
\cite{Callan:1994ub}. Then, as far as boundary states are concerned,
one can raise the value of the tachyon field ``past infinity'' to a
whole different sector not available classically. In particular, a
$\sutwo$ rotation will bring the original Neumann state to a different
one: 
\be 
e^{2\pi i J_1}|N\rangle_{\tiny
  \sutwo}=2^{-\frac{1}{4}}\sum_{j,m} e^{2\pi i j}
|j,m,-m\rangle\rangle,\label{neumannishi} 
\ee 
which still satisfies the Neumann condition in the absence of Wilson
lines $(J_3+\bar{J}_3)|N\rangle=0$. There is a similar ``doubling'' of
the Dirichlet state. These differentiations can be modelled in the
proposed association by seeing the $\sutwo$ element as an ordinary
unitary $2\times 2$ matrix. The group geometry allows one
to see that the region covered by classical values of the tachyon
field is only a coordinate patch of the full space of boundary
perturbations.  This feature is particular to the global $\sutwo$
structure that shows up at the self-dual radius. In fact, it is the
existence of Ishibashii states with half-integer $j$ in
(\ref{neumannishi}) that makes for the non-triviality of the $2\pi$
rotation. By contrast, in type 0B string theory where one has instead
the ``real'' ${\rm SO(3)}$ structure, there will be no doubling.

One can extend these two cases to the Lorentzian theory by performing
a Wick rotation to the generator $J_3$, bringing the group structure
to $\sltwor$. The generator $J_1$ will now yield the elliptic
subgroup of $\sltwor$, ${\rm U(1)}$. To make the correspondence
precise, we will take the Dirichlet state as a reference, where the
group element associated to it will be the identity. Fixing $\alpha=0$ and
performing the Wick rotation $\phi=it$ in (\ref{euler}), one has
\be 
g=e^{-t  J_3}e^{2i\theta J_1}e^{t J_3},
\ee 
whose values of $t$ and $\theta$ parametrize the allowed, physically
distinct one-brane configuration states. The $J_3$ current now is
interpreted as the Hamiltonian, implementing $t$ translations, and
$\theta$ is related to the value of the tachyon, in a manner that
$\theta=\pm \pi/2$ correspond to the Neumann states. Note that the
latter are invariant by $t$ translations, and then fixed points of the
condensation process. 

The condensation process is then viewed as the flow of $J_3$ in this
configuration state. The geometrical action for this flow for a curve
parametrized by $s$ is 
\be
S_{\tiny\rm single}=-m\int ds \sqrt{-{\rm Tr}[(g^{-1}\dot{g})^2]}=-m\int
ds~\sqrt{-\dot{\theta}^2+ \cos^2\theta\,\dot{t}^2}, \label{dbi}
\ee
which, upon the change of coordinates $\cosh\tau=\sec\theta$ can be
recognized as the DBI action, the worldline action for a single
D-brane. Note that only the sector between $\theta=-\pi/2$
(corresponding to $|N'\rangle$) and $\theta=\pi/2$ ($|N\rangle$) is
mapped through this transformation. Classically this poses no problem
since the movement for finite $t$ is restricted to this region. 

\begin{figure}[hbt]
\begin{center}
\includegraphics[height=5cm]{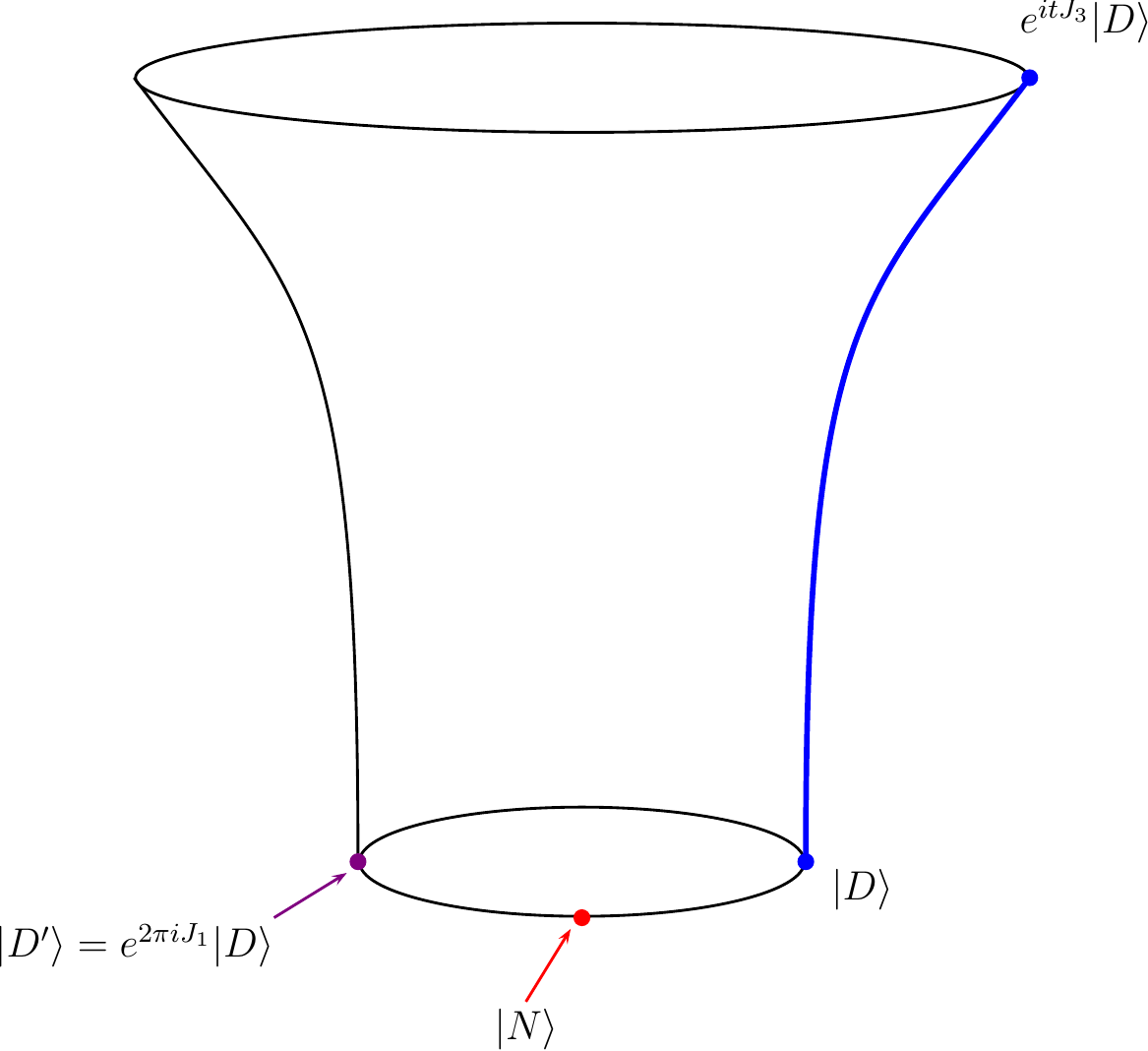}
\end{center}
\caption{\em The global geometry of the configuration space for a
  single brane, in the case where one has a Wick-rotated counterpart of
  the self-dual radius. One needs to act by $e^{4\pi J_1}$ for a full
  turn.}\label{hyper}
\end{figure}

The geometry in itself is quite interesting. The metric that arises from
the parametrization above for the Euclidean case is that of a sphere
$d\theta^2+\cos^2\theta\,d\phi^2$, whereas in the Lorentz
case we have a two-dimensional space with constant curvature: 
\be
ds^2=d\theta^2-\cos^2\theta\, dt^2 \label{metric} 
\ee 
which can be
represented isometrically as the one-leaf hyperboloid in
$\mathbb{R}^{2,1}$ (Fig. \ref{hyper}.) 

In the Euclidian theory, one can interpret the ``new'' Dirichlet as
being the antipode point to $|D\rangle$ in the sense that the former
is labelled by $\theta=0$ whereas the latter has $\theta=\pi$. In the
Lorentzian theory this still holds, but one can define an ``imaginary
position'' to $|D'\rangle$ via the metric (\ref{metric}). From the
metric one finds the $\sltwor$-invariant distance between two points
$p$ and $p'$: 
\be
Z(p,p')=\cosh d(p,p') = \cosh(t-t')\cos\theta\cos\theta'-
\sin\theta\sin\theta'\label{distance}
\ee
from which one sees that $|D'\rangle$ sits at the ``imaginary
distance'' $d=i\pi$. Note that $d(p,p')$ is periodic in $\theta$, and
this periodicity will translate to an ``imaginary periodicity'' in $t$
when one computes $\sltwor$ invariant quantities as we will do in the
next session. For instance, an $\sltwor$-invariant propagator will
display this periodicity in imaginary time, a thermal behavior. For
type 0B the same argument as above yields half of the periodicity, as
if $|D\rangle$ and $|D'\rangle$ were identified.  The periodicity must
be mimicked in models of tachyon condensation without the geometric
picture. Tachyon insertions must be replaced by an array of insertions
at imaginary positions to yield the closed string amplitudes
\cite{Gaiotto:2003rm}. 


One notes then that the $\sltwor$ structure
allows one to relate the radius of compactification in the
Euclidianized theory to the radius of compactification of the variable
$\theta$ above. Because of the local $\sltwor$ structure, picking a
different compactification of $\theta$ amounts to truncating the action
of the universal covering group of the local structure,
$\widetilde{\sltwor}$. One is thus tempted to argue that the quantum
mechanical states in generic tachyon condensation process can be
obtained from some suitable truncation of the universal covering group
$\widetilde{\sltwor}$. We will give some support for this argument in
the next section.

\section{Dense Packing of Branes}

In the Euclidean theory, the fact that the state is on-shell means
that the quantities $\theta_a$ are constant throughout the
renormalization group flow. The analogy with movements in space is
clear at zero coupling: being heavy particles $m\propto 1/g$, the D0
branes are well-localized in the configuration space. In real time
formalism, time evolution is implemented by the action of $J_3$. The
condensation process is then the change of the ``coordinates'', or
the values of fields, by the action of the time evolution operator. The
Lagrangian obtained from this symmetry is the DBI action (\ref{dbi}).

One would expect that the configuration space of a large number of
branes at finite coupling to be completely different from the one
described in the preceeding section, but surprisingly this is not the
case. In the spirit of the preceeding section, let us associate a
position in the space of allowed boundary conditions to each brane,
that is, a $g_i\in\sltwor$. The action of $N$ of these branes will
have a natural expansion in terms of $N$ particle interactions: 
\be
S(\{g_i\})=\sum_{i\neq j}\lambda_2 S_2(g_i,g_j)+\lambda_3 \sum_{i\neq
  j\neq k\neq i} S_3(g_i,g_j,g_k)+\ldots\label{mostgeneral} 
\ee 
where the ${\rm U(N)}$ symmetry act as shuffling the indices. As such,
the system has, along with the shuffling symmetry, the global
$\sltwor\times\sltwor$ symmetry $g_i\goesto gg_ih$. The $\lambda_i$
are coupling constants, which scale with the open string coupling
$\lambda_n\propto g_o^n$. One can think of the $S_2$ term as measuring
the ``distance'' in moduli space between brane ``i'' and ``j''. Note
that now there is no {\it a priori} concept of distance between the
indices, as opposed to usual matrix models. The action above is of
course too general to be of any use but it can be reduced to an
``universal form'' through a few extra assumptions, or restrictions,
on which brane configurations are considered. 

The first and most important one is the assumption that the packing of
branes will be {\it dense}, {\it i.e.}, that we can label the branes
by indices $i$ which take continuous values with the local topology of
the target space, that is to say, $\mathbb{R}^3$. This is different
from the usual matrix models in which the indices have the topology of
the real line locally. Choosing this topology of $\mathbb{R}^3$ is
natural from the moduli space point of view. Over these
configurations, one can give a simple argument to symplify the
intractable structure of (\ref{mostgeneral}). The relevant 2-point
function has the structure:
\be
\langle g_i | g_j \rangle_{\lambda} = D(g^{-1}_ig_j) + \lambda
\sum_{g_k}D(g^{-1}_ig_k)D(g_k^{-1}g_j)+\ldots =
D_\lambda(g_i,g_j), 
\ee
where $D(g)$ is the matrix representing the action of the group in the
Ishibashii states. The sum will of course turn into an integral over
all the intermediate configurations, but the relevant point is that
whichever is the result of the integral, it will also be a class
function of $g_i$ and $g_j$, that is, it will be a function of
$g_i^{-1}g_j$ of the trace type,
$D_\lambda(g_i,g_j)=D_\lambda(g_i^{-1}g_j)$. Such functions can be
written as a sum of representations of the group, and as such are just
redefinitions of $D(g)\goesto D_\lambda(g)$. Since the particular
representations which enter into $D(g)$ are not relevant to classical
dynamics of the $g_i(s)$, we can then consider the first term only.

One has to be a bit more careful with the last statement. It is true
that the form of the matrix $D_\lambda(g)$ does not matter for the
dynamics of the zero modes, but one has to be careful to check whether
non-unitary representations will be included in the expansion
above. Like in the last section, we take the view of 
the tachyon condensation process as the dynamics of boundary
states. In general, the latter are the Ishibashii states, which are
superpositions of highest weight states of all the unitary
representations of the current algebra. One is expected, then, to have
only unitary representations in the dynamics of the zero modes. This
is extremely important since in the Lorentzian case the symmetry group
is non-compact and hence unitarity could be broken. By this argument
this situation will not arise.

Along with the density of the configurations, the other ingredient
which will be relevant to this discussion is a well known result from
matrix models. The large number of particles leads us to consider
non-abelian tachyon fields, and the singlet sector of this field under
the symmetry of exhange of indices can be represented by Fermi
statistics on the single particle configuration space. One can then
introduce a spinor operator $\psi(x_i)$ which populates an eigenvalue
(D0-brane, or boundary term) of the particle $i$. The singlet sector
will then be the ``vacuum'' state of this field operator, over which
we will perturb the dynamics. Given that unitarity is preserved,
spin-statistics tells us that this field should transform under the
spinor representation of $\sltwor$. 

With the discussion above one sees that the type of correlation
function we are interested are of the type:
\be
\langle \psi(x_i)g(x_i)^{-1}g(x_j)\psi(x_j)\rangle
\ee
where $\psi(x_i)$ is the Fermi field that parametrizes the
eigenvalues' distribution. The matrix $g(x_i)^{-1}g(x_j)$ parametrizes the
overlap between the boundary states of the ${\rm i^{th}}$ and ${\rm
  j^{th}}$ eigenvalues. Now one sees why the details of the
representations which take part of $D_\lambda(g)$ are not important:
only the boundary state representative $g$ and its action on the
spinor, which must transform in the fundamental of $\sltwor$ matter
for the discussion. The effect of turning the coupling will be to
change the fundamental representation to an equivalent one.  We will
omit from now on the indices $i$ and $j$ and assume a continuous
distribution parametrized by three coordinates $x$. The argument above
is reminiscent of the sewing techniques used in String Field Theory
\cite{DiVecchia:1989ht,Horowitz:1987rm}. 

We must also remember that the global configuration of the boundary
states depends on the particular model of tachyon condensation we are
taking. In the Euclidian case, for bosonic strings compactified at the
self-dual radius this global structure is, $\sutwo$. For type 0B it
is ${\rm SO(3)\approx SU(2)/\mathbb{Z}_2}$. As we argued in the last
section, in general it will be some subgroup of the covering group of
$\sltwoc$. In order to avoid these global complications, we will
consider the invariant action based on a local connection of the gauge
group: 
\be 
S[\psi,\bar{\psi},A_\mu] = k^\prime \int
d^3x\,\bar{\psi}\gamma^\mu(\partial_\mu+A_\mu)\psi\label{baseaction},
\ee 
where the condition for the existence of the map $g(x)$ which
gives the boundary perturbation for each D-brane is translated to the
flatness of the ${\rm sl(2,\mathbb{R})}$ connection: 
\be
\partial_{[\mu} A_{\nu]}+A_{[\mu}A_{\nu]}=0. \label{flatconnection}
\ee

The solutions of (\ref{baseaction}) with flat connection are spinors
of the type $\psi(x)=g(x)\eta$ with $\eta$ constant transforming in
the fundamental of $\sltwor$. These configurations do indeed
correspond to ``occupied'' localized states for all values of $x$. The
extra terms -- like gamma matrices -- in the action are defined so
that (\ref{baseaction}) is invariant under the choice of the
Killing-Cartan form we pick for $\sltwor$. In fact in this case a
particular choice of the gamma matrices is nothing more than a choice
for a basis of the Lie algebra. It will also be in our interest to
keep the connection $A_\mu$ arbitrary, since the requirement
(\ref{flatconnection}) will turn out to be redundant.

If we integrate the fermions the resulting effective action for
(\ref{baseaction}) will be the usual Chern-Simons, written in terms of
the 1-form $A=A_\mu \,dx^\mu$:
\be
S_{\rm eff}=\frac{k}{4\pi}\int {\rm Tr}\left(A\wedge
  dA+\frac{2}{3}A\wedge A\wedge A\right), \label{chern-simons}
\ee
which represents the fact that the coordinates are just dumb indices
whose metric structure cannot influence the dynamics. As promised
above, the condition (\ref{flatconnection}) is recovered by the
equations of motion of (\ref{chern-simons}). One can see the action
above as a result of quantum fluctuations of the $\psi$ field around
the ``vacuum'' (singlet) state. In this case the 0-branes coalesce and
a space-time, closed string metric structure arise under certain
limits, as we will see in the remaining of the paper.

\subsection{Of Gauge Choices and Constraints}

The Chern-Simons term which dictates the dynamics of the map $g(x)$
has, as it is well-known, no local degrees of freedom. Non-trivial
dynamics will arise if the domain of the map $g(x)$ has
boundaries. This translates to the physical condition that our
D0-branes cover only a finite-volume subset of the target-space. The
reduction from (\ref{chern-simons}) to the boundary is well
studied (see, for instance, \cite{Alekseev:1989npb}). The effective
action will be: 
\be
S=\frac{k}{2\pi}\int d^2z ~\mbox{Tr}(g^{-1}\partial g
g^{-1}\bar{\partial} g)+\mbox{constraint},
\label{WZW} 
\ee
where the constraint's purpose is to gauge the symmetry
(\ref{symmetry2}). The reason behind the promotion of
(\ref{symmetry2}) from a global to a local symmetry is that one cannot
physically control the value of the axial coordinate $\alpha$ chosen
for all the D0-branes. By gauging the axial action, one removes the
redundancy on the value of the tachyon field introduced by the
boundary state group coordinates.

Let us further clarify the the form of the constraint. For
the movement of a single particle, its effect is
just the introduction of the Lagrange multipliers in the geometrical
action:
\be
S_{\rm single}=\int
ds~\mbox{Tr}(g^{-1}\dot{g})^2+\lambda(s)\dot{\alpha}(s) 
\label{actionsingle},
\ee
with $g$ and $\alpha$ as in (\ref{euler}). One can alternatively think
of the extra term as arising from the imposition of local invariance:
\be
g(s)\goesto h(s)g(s)h(s),
\ee
with $h(s)$ as in (\ref{symmetry2}), which has the advantage of not
choosing coordinates {\it a priori}. This reproduces the constrained
Lagrangean after one substitutes the equations of motion for the gauge
field $A = h^{-1}\dot{h}$, or integrate it out in the path integral
formalism. For the fluid, we would rather use the second
formalism. If the real parameter $u$ labels the branes at the
boundary, one requires on (\ref{WZW}) the invariance:
\be
g(s,u)\goesto h(s,u)gh(s,u),
\ee
which introduces the following term in (\ref{WZW}):
\be
\mbox{constraint}=
\mbox{Tr}(g^{-1}\partial{g} A_{\bar{z}})+
\mbox{Tr}(\partial{g}g^{-1} A_{\bar{z}})+{\rm c.c.}+
\mbox{Tr}(A_zA_{\bar{z}}) .
\ee

To better connect this proposal with the literature, one can think of
the action (\ref{WZW}) as the bosonization of the DBI matrix model
introduced in \cite{daCunha:2004dh}, or simply the effective action
one has when the collective field derived from (\ref{dbi}) is coupled
to $\sltwor$ currents.  
\be
T^i_a=\frac{i}{2}\left[\bar{\psi}\gamma^i\nabla_a\psi-
  \nabla_a\bar{\psi}\gamma^i\psi\right],
\ee
where $i$ runs through the $\sltwor$ indices. The equations of motion
of (\ref{WZW}) will then enforce the Ward identites of the collective
field currents. 

Fixing the coupling -- or, alternatively, the mass of the brane --
amounts to fix the value of the quadratic Casimir $J^2$. In the model
above, this means that the three components of the currents will not
be independent, since their sum of squares add to the value of the
Casimir $m^2=\frac{1}{4}+\mu^2$. Also, in order to rewrite the matrix
model as the fermionic field, one has to impose the Gauss constraint
of the gauge field. However, for finite $k$, this is not quite the
exact thing to do \cite{Gawedzki:1988nj}: in order to deal with the
gauge symmetry one has to introduce ghosts and deal with the BRST
quantization of (\ref{WZW}). Luckily this particular program has been
accomplished some time ago and we are left with the job of
reinterpreting the spectrum.

\subsection{$\sltwor$ representations and branes.}

Before analyzing the closed string excitations and the corresponding
formulation in terms of the Liouville field, let us digress over the
excitations of the theory above. The spectrum of the Euclidean theory
will be then dictated by the representations of $\sltwor$. In 
this case these are labelled by the quantum numbers
$|j,m\rangle$, $j$ a positive half integer and $|m|\leq j$, and whose
wavefunctions are spherical harmonics in the $\theta, t$ plane. These
correspond to the usual ``special states'' of the two dimensional string,
{\it sans} the appropriate Liouville dressing, and an action
equivalent to (\ref{chern-simons}) has been proposed by Klebanov {\it et
  al.} \cite{Klebanov:1991hx}. We stress the appearance of the spinor
degrees of freedom: from the fact that a single boundary perturbation
transforms in the fundamental representation, one can associate the
two independent solutions for the tachyon profile $e^{\pm i
  X(0)/\sqrt{2}}$ as the ``spin up'' and ``spin down'' states. One
must also note that this analysis is again valid only insofar the
``gauge fixing'' $J_2=0$ can be consistently made, {\it   i. e.},  in the
large $k$ limit. In the generic case one has to dress 
the states with the Liouville field as in the construction of the
reference above. It is found that only the states with $m=j-1$ survive
the gauging process.

In the Lorentzian case, the situation changes somewhat.  The sectors
$J^2=-(\frac{1}{4}+\mu^2)$ span the principal continuous series of
$\sltwor$.  If we assume the compactified orbits of the elliptic
generator $J_3$, the spectrum of each sector consists of states
\be
{\cal H}=\left\{|j=-\half+i\mu, m+\nu\rangle\right\},\quadsp m\in
\mathbb{Z},\,\,\nu\in [-\half,\half)\label{hilbert}
\ee
One can think of the phase $\nu$ as parametrizing the twist of the
boundary conditions on the matrix model -- and therefore on the
collective field. In \cite{Kazakov:2000pm} these where introduced to
study the non-singlet sectors. From the open string perspective one
would expect these to arise from compactification of the Euclidian
boson (or array of branes at ``imaginary positions'') to non-rational
radii (separation), and/or the turning of fluxes. For now let us
consider the identity sector $\nu=0$.

First let us consider the density of levels. The actual number of
levels is of course actually infinite:
\be
{\cal N}(\mu)=\sum_{m\in\mathbb{Z}}\langle j,m| j,m\rangle
\ee
but the sum can be regularized in an invariant way. For this consider
the function
\be
{\cal G}(\mu;t,\theta)=\sum_{m\in\mathbb{Z}} \langle j,m|
D(g(t,\theta)) |j, m\rangle
\ee
where $g(t,\theta)$ is as in (\ref{euler}) and $D(g)$ is the
representation of the $\sltwor$ matrix $g$. One can then see that the
character ${\cal G}(\mu;t,\theta)$ is a Green's function of the scalar
Laplacian on the metric (\ref{metric}), and then it can be interpreted as a
two-point function of a scalar field in that space, whose mass is
given by $M^2=J^2=\frac{1}{4}+\mu^2$. The two-point function at
coincident points can be regularized by usual means \cite{Candelas:du}
to yield:
\be
\varrho(\mu)=\frac{1}{2\pi}
\left(-\log\Lambda+\psi(-\half+i\mu)+\psi(-\half-i\mu)\right)
\label{density}
\ee
where $\psi(x)$ is the Euler digamma function. Comparing the
expression above with the usual Matrix Model calculations, one finds
readily that the UV cutoff introduced $\Lambda$ has the interpretation
as the {\em number} of branes (or the size of the matrices). In this
formulation, the large $N$ limit is exactly the same as the UV
limit. Also, the parameter $\mu$ has the interpretation of the double
scaled inverse coupling, and the expansion of (\ref{density}) in
inverse powers of $\mu$ show the characteristic $1/(2n)!$ behavior
of closed strings \cite{Shenker:1990uf}. It is interesting to point
out the embedding of such field in the fermionic collective field
model, by making use of the identity involving Jacobi functions
\be
\cosh\frac{d}{2}\, {\mathscr D}^{-\tinyhalf
  +i\mu}_{-\tinyhalf,-\tinyhalf}(\cosh d)+\sinh\frac{d}{2}\, {\mathscr
  D}^{-\tinyhalf +i\mu}_{-\tinyhalf,\tinyhalf}(\cosh 
d)={\mathscr D}^{-\tinyhalf+i\mu}_{0,0}(\cosh d).
\ee
The left hand side of the equation above can be seen to be the two
spinorial components of the Green's function of a spinor field in the
metric (\ref{metric}), whereas the left hand side is proportional to
${\cal G}(\mu;t,\theta)$. One can see that the bosonic excitations are
seen as superpositions of two fermionic ones, with these two related
by discrete symmetries. The same remarks where made in
\cite{Douglas:2003up}, although this makes clear that the end result
of the condensation process should be stable, and the apparent
instability of bosonic string should be an artifact of an artificial
truncation of the spectrum by, for instance, ``chopping off'' the
excitations with $\tau<0$ and hoping that tunneling through the
barrier will not happen. 

\begin{figure}[htb]
\begin{center}
 \includegraphics[height=4cm]{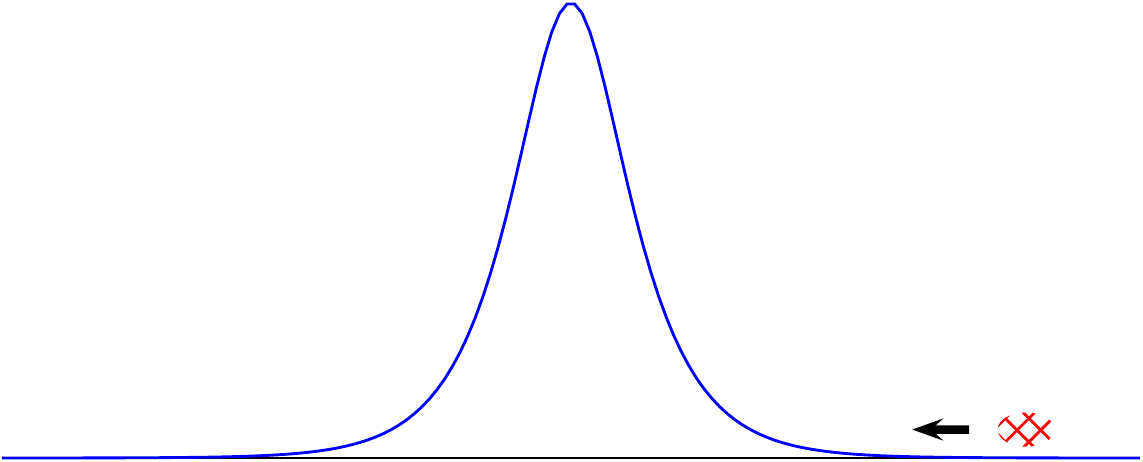}
\end{center}
\caption{\em A pictorial representation of a D0-brane type of
  excitation in the $\tau,t$ plane. The effective classical
  potential $M/\cosh\tau$ is also  drawn.}\label{particle} 
\end{figure}

This relationship will also be useful to gain insight on the
model. Upon change of coordinates $\cosh\tau=1/\cos\theta$ the
equation for ${\cal G}(\mu;t,\tau)$ becomes
\be
\left(\frac{\partial^2}{\partial t^2}-\frac{\partial^2}{\partial
    \tau^2}+\frac{M^2}{\cosh^2\tau}\right) {\cal
  G}=\delta(t)\delta(\tau),\label{klein-gordon}
\ee 
with $M=-J^2=\frac{1}{4}+\mu^2$. So one can consider for the purposes
of illustration an effective classical motion given by the
(``tachyon'') potential $V(\tau)=M/\cosh\tau$. In order to recover the
classical limit one has to construct coherent states out of the
spectrum. The starting point to this effect is the observation that we
can use the states (\ref{hilbert}) to construct states localized in
$\theta$, or equivalently, localized in the tachyon value: 
\be
|j,\theta\rangle = \sum_{m\in\mathbb{Z}}e^{-im\theta}|j,m\rangle ,
\ee
in which $j=-\half+i\mu$ is a constant. These states span all the
states available for a fluid moving in the metric (\ref{metric}). With these in
mind one can construct delta-function normalized eigenstates
of the Hamiltonian $J_3$, which is done in \cite{Barut1965,Vilenkin}: 
\be
|j,\omega\rangle_+=\frac{1}{\sqrt{2\pi}}\int_{-\frac{\pi}{2}}^{\frac{\pi}{2}}
d\theta~ \sqrt{1+\sin\theta}^{\,j-i\omega}
\sqrt{1-\sin\theta}^{\,j+i\omega}|j,\theta\rangle .\label{eigenj3}
\ee
Now we would like to draw the readers' attention to the
``wavefunction''
\be
\phi_\omega(\theta)=\frac{1}{\sqrt{2\pi}}\sqrt{1+\sin\theta}^{\,j-i\omega}
\sqrt{1-\sin\theta}^{\,j+i\omega}
\ee
on which we would like to make two comments. Firstly, the probability
distribution is {\em independent} of $\omega$,
$|\phi_\omega(\theta)|^2=(\cos\theta)^{-1}$. By the change of
variables $\cosh\tau=(\cos\theta)^{-1}$ one arrives at an expression
equivalent to (\ref{eigenj3}):
\be
|j,\omega\rangle_+=\frac{1}{\sqrt{2\pi}}\int_{-\infty}^{\infty}d\tau~
e^{i\omega\tau}|j,\tau\rangle_+,\label{eigenj3tau}
\ee
in which one redefines $|j,\tau\rangle$ to absorb the $\tau$-dependent
phase -- $(\cosh\tau)^{-i\mu}$ -- coming from the integral. Then one
can represent a localized excitation (a D0 brane) as a minimum uncertainty 
wavepacket in the $t,\tau$ plane as represented in Fig. (\ref{particle}).

The second comment on (\ref{eigenj3}) is about the fact that one does
not need all of $|j,\theta\rangle$ to construct an eigenstate of
$J_3$. In fact, only ``half'' of the circle suffices. This ``double
degeneracy'' is a
characteristic of the representation theory of $\sltwor$ (see, for instance,
\cite{Kuriyan1968}) and stems from the fact that the action of $J_3$ on
the configuration space has two fixed points -- the states $|N\rangle$ and
$|N'\rangle$ in the language of section 2. By the action of $J_3$
states with $|\theta|<\pi/2$ are brought to states with
$|\theta|<\pi/2$, and likewise for the other side of the circle,
justifying the suffix ``$+$'' in (\ref{eigenj3}) and in
(\ref{eigenj3tau}). One can at this point posits that, for the case of
a configuration space given by $\widetilde{\sltwor}$, one has in fact
to include an infinite number of these sectors. For our purposes, a
single sector will suffice, since the dynamics of a single brane
excitation is confined to it. 

\section{Closed string formalism}

It is well-known that the action (\ref{WZW}) has a direct
interpretation in terms of closed strings moving in a particular
background \cite{Dijkgraaf:1991ba}. It is however less direct to see
how exactly the closed string excitations result from the tachyon
field.  As stated above, the usual collective field formalism for the
tachyon condensation process involves a gauge-fixing which is not
exactly natural from the closed string point of view. They should be
nonetheless equivalent as long as the number of particles $N$ is
large, and thus we barge ahead and try to model Liouville like
excitations from the collective field formalism.  To add to the
confusing literature, we present two distinct ways of doing so.


\subsection{The Non-relativistic (McGreevy-Verlinde) limit}

As we saw in the preceeding section, after the gauge fixing, the
tachyon process can be modelled by a relativistic particle moving in a
$\sech$ potential. Since the $c=1$ matrix model involves a inverted
harmonic oscillator potential, it is natural to think that the double
scaling limit will single out the maximum as the region of
interest. As a matter of fact, the expansion of the potential $M/\cosh
\tau$ will indeed yield an inverted harmonic potential. Borrowing from
the literature, one considers the generating function for tachyon
expectation values: 
\be
W(q,t)=\langle\Psi |e^{q J_1}|\Psi\rangle, \label{sou}
\ee
more specifically, variations of $W$ around states with definite energy
$J_3\approx\mu$. The quantity $W$ is the quantum analogue of the
inverted harmonic oscillator quantity given by:
\be
W(\ell,t)\approx\int d\tau e^{\ell\tau}(p_+(\tau)-p_-(\tau)),\label{sou2}
\ee
considered in the semiclassical limit, where expectation values are
substituted by integrations over the Fermi sea. The boundary of the
Fermi sea has momenta $p_\pm(\tau)=\pm\sqrt{2M\mu+M^2\tau^2/2}$, 
which enter into the semiclassical expression (\ref{sou2}).
The quantity $W$ gives the ``size of the universe'' observable in Liouville
theory. Since the time translation generator in (\ref{sou}) is given
by $J_3$, one can compute the second derivative of $W$ as in
(\ref{sou}) with respect to time by using the identity:
\be
[[e^{q
  J_1},J_3],J_3]=(-J_1\sinh q+J_2^2\sinh^2q-
\{J_2,J_3\}\sinh q(\cosh q-1)+ J_3^2(\cosh q -1)^2)e^{q J_1}, 
\ee
Of which we perform a contraction: 
\be
J_{1,2}\goesto \sqrt{\rho}L_{1,2}, \quadsp q=\ell/\sqrt{\rho} 
\mbox{ and }\rho\goesto \infty.\label{contraction}
\ee 
Because $|\Psi\rangle$ has a definite value for  $J^2$, one
can then compute ${J_2}^2$ in terms of $\mu$,
$q$ and $J_3$, the time derivative operator. Only the 
first and the second terms on the right hand side survive the
contraction, and the result is
\be
\frac{\partial^2}{\partial t^2}W=
\ell\frac{\partial}{\partial \ell} W+
\ell^2\frac{\partial^2}{\partial \ell^2} W-
((J_3)^2-\mu^2)\frac{\ell^2}{\rho} W+{\cal O}(\rho^{-1})\label{wdw1}
\ee
so, by substituting $J_3=\mu-H$, with $H\ll \mu$ and taking
$\mu=\rho$ one accomplishes the non-relativistic
limit. The Wheeler-de Wit equation arises as one deforms the states
$|\Psi\rangle$ by, say,  adding an eigenstate of $H$
\cite{Ginsparg:1993is}. 

In this picture the positive and negative energy solutions, as
measured by the sign of $H$, give rise to opposing signs in last
relevant term in (\ref{wdw1}). However, since we consider the Fermi
sea to be filled up to $J_3\approx\mu$, the right excitation is the
absence of a negative energy state, which has the ``right'' sign for
the usual correspondence between the Wheeler-de Witt equation and the
Liouville field with a negative cosmological constant
term \cite{Ginsparg:1993is}. The ``hole'' states thus give 
rise to another closed string sector described by another Liouville
field which is independent in the weakly coupled regime $\mu\goesto
\infty$. It would be interesting to go one order further in $\mu^{-1}$
to compute the mixing between the two sectors, but we will leave such
discussion to future work.

\subsubsection{Double scaling limit}

It is perhaps worth pointing here that it is not clear that the double
scaling limit will actually perform the process of contraction
referred above. In fact, the computation of the density of levels done
in the last section hints strongly that the couplings in (\ref{WZW})
-- particularly $\mu$ -- are already scaled and hence there is no
physical ground to the substitution of variables done in
(\ref{contraction}) other that it accomplishes the non-relativistic
limit of the potential (\ref{klein-gordon}).  Another argument against
this comes from the density of levels of the flux backgrounds
\cite{Maldacena:2005he}. By changing the value of $\nu$ one selects
amongst representations of $\sltwor$. The density of levels is
computed using the regularization of the character of the identity as
above (\ref{density}). The relevant matrix element involves the Jacobi
functions \cite{Vilenkin}: 
\be 
\varrho(\mu,\nu)=N_\nu\lim_{Z\goesto 1}
{\mathscr D}^{-\tinyhalf+i\mu}_{\nu,\nu}(-Z)=\tilde{N}_\nu
~{_2F_1}(\half-i\mu, \half+2\nu+i\mu;1+\nu,\half (1-Z)) 
\ee 
where $Z$ is as in \cite{daCunha:2004dh} and ${_2F_1}$ is the usual
hypergeometric function. Expanding the expression around $Z=1$ one
finds the known expression involving the digamma function:
\be
2\pi \varrho(\mu,\nu)=\frac{1}{\epsilon}+\Re\psi(\nu-\half+i\mu) 
\ee 
which is exactly the one found in \cite{Maldacena:2005he}. This infers
that those representations with $\nu\neq 0$ model flux backgrounds,
with the flux given by $\nu$.

If the non-relativistic limit is to come from a sensible open-string
picture, one has to compactify the tachyon direction to 
make the energy finite, as in \cite{McGreevy:2003kb}. One notes that
such identification is done with respect to the tachyon
value, here referred to as $\tau$, and not the affine parameter of the
Killing vector field, or the \sltwor isometry operator $J_1$, which is
here called $\theta$. Thus such identification is unnatural from the $
\sltwor$ perspective.

\subsection{The ultra-relativistic (Sen) limit}

The other way of constructing Liouville-like excitations is the well
known coadjoint orbit reduction \cite{Alekseev:1989npb} of the \sltwor
model. This can be understood geometrically as follows. Consider a
particle sitting at the top of the potential (see Figure
\ref{particle}).  Upon an \sltwor rotation, the time translation
operator $J_3$ is transformed to: 
\be e^{-i\theta J_1}J_3e^{i\theta
  J_1}=\cosh\theta J_3+i\sinh\theta J_2.  
\ee 
Now, as one takes the limit $\theta\goesto\infty$, the time
translation symmetry is transformed to $J_+$, the generator of the
elliptic subgroup of \sltwor. By attributing a determined value for
the ``energy'', {\it i. e.}, by restricting ourselves to the sector
$J_+=\sqrt{\mu}$, one accomplishes the reduction from the \sltwor
model to Liouville \cite{Dijkgraaf:1991ba}. One can understand the
``boost'' made above as the ``pushing'' of the Fermi sea to infinity
in Figure \ref{particle}. According to Sen, that is the minimum of the
tachyon potential and where the closed string excitations are supposed
to be localized. These excitations are exactly the ``near-horizon''
modes alluded to in \cite{daCunha:2004dh}.

Such excitations are modelled by states of constant $j=-\half+i\mu$ in
the ``classical limit''. Their dynamics can be recast in the form of a
scalar field in a curved space, whose geometry is given by the
geometry of the coset space (\ref{metric}). This is two-dimensional de
Sitter, and hence all physical properties of the condensation process,
and of the closed string excitations, can be interpreted as properties of the
propagating field in de Sitter. The first unsettling connection one
makes is that the ``black-hole'' state in the coset model (of closed
strings) has the interpretation of a ``de Sitter'' ({\it i. e.},
$\sltwor$-invariant) vacuum state in the collective field of open
strings.  This shows an equivalence, at least in the two-dimensional
model, between the black hole horizon and the cosmological horizon,
the novelty being that the equivalence stems from the old open-closed
string duality. 

\section{Discussion}

In this article we discussed some semiclassical points pertaining to
the condensation of D0 branes in non-critical string theory. It was
argued how the spectrum of excitations should fit in the WZW coset
model, with states in the former filling the representations of
$\sltwor$. We gave the interpretation of the condensation as the
collective movement on the moduli space of allowed boundary states,
and discussed two distinct limits where the closed string spectrum is
recovered. Finally one posits that the black hole state constructed in
the closed string sector should in fact represent a cosmological state
constructed in the open string collective field.

One should note that this ``geometrization'' of the view of the
condensation process is in fact necessary to avoid confusing
coordinate transformations in field which only work in a patch of the
configuration space. On the other hand true geometrization can only
be achieved in the semiclassical limit and thus we will refrain from
proposing a strict equivalence between the different points of
view. One should remark, however, that the old mantra that all allowed
boundary states correspond to closed string excitations is not
followed in this work: the open string states corresponding to fluxes
are gauge degrees of freedom and hence not dynamic. These states have
to be dealt with by a gauge choice, which arises the question whether
a global gauge fixing choice can be achieved. While the Coulomb gauge
is usually used to yield Liouville theory in the double scaling limit,
one is also tempted to not fix the gauge straight away and arrive
through the symmetry argument presented here to the coset model. In
the spirit of ``covariance'' the second method should be preferred.

What is true, nevertheless, is that one may see the spectrum of closed
strings from the usual matrix model and the ``near horizon'' spectrum
obtained here as different entities, arising from different limits,
and gauge choices, from the picture of the dynamical variables in the
boundary state formalism. One should note that the latter proposal
implements Sen's ideas of how the closed degrees of freedom are
encoded in the open string formalism and need no artificial
constructions like an external thermal bath to keep the energy
$\langle J_3\rangle$ close to $\mu$ like in (\ref{wdw1}).  

One is left wondering on what lessons can be transported to more
realistic scenarios of tachyon condensation. Despite being a difficult
guess to hazard, there is one redeeming feature of the study conducted
here: the major pillar of sustentation of the work is the open-closed
string duality. In our particular case this comes about as the
T-duality that relates on the local group structure, Euclidianized to
$\sltwor$. Whether there are topological issues other than fluxes and
broken symmetries that arise in the general problem is an open problem
we would hope to return in the future.

\section*{Acknowledgments}
To Lucas Alves Carneiro da Cunha, with whom some conversations around
the main topic of section 3 proved useful. I would also like to thank
Shinsuke Kawai, Esko Keski-Vakkuri, Vasilis Niarchos and Mischa
Sall\'e for comments and suggestions. I wholeheartedly thank the
Helsinki Institute of Physics for support during most of this
work. The author would like to apologize in advance for any omissions.

\end{document}